\documentclass[showpacs,prl,twocolumn]{revtex4}
\pdfoutput=1
\usepackage{textcomp}
\usepackage{amsmath}
\usepackage{amssymb}
\usepackage{graphicx} 
\usepackage{pgf}
\newcommand{\ket}[1]{\left|#1\right\rangle}
\newcommand{\bra}[1]{\left\langle #1\right|}
\newcommand{\ud}{\mathrm{d}}

\newcommand{\nep}{\textrm{e}}

\begin{document}

%
\title{Periodic steady regime and interference in a periodically driven quantum system}

\author{Angelo Russomanno$^{1,2}$, Alessandro Silva$^{3}$, Giuseppe E. Santoro$^{1,2,3}$}
\affiliation{
$^1$ SISSA, Via Bonomea 265, I-34136 Trieste, Italy \\
$^2$ CNR-IOM Democritos National Simulation Center, Via Bonomea 265, I-34136 Trieste, Italy \\
$^3$ International Centre for Theoretical Physics (ICTP), P.O.Box 586, I-34014 Trieste, Italy
}

\begin{abstract}

We study the coherent dynamics of a quantum many-body system subject to a time-periodic driving. 
We argue that in many cases, destructive interference in time makes most of the quantum averages time-periodic, after an initial transient. 
%
We discuss in detail the case of a quantum Ising chain periodically driven across the critical point,
finding that, as a result of quantum coherence, the system never reaches an infinite temperature state.
Floquet resonance effects are moreover observed in the frequency dependence of the various observables, which display a sequence of well-defined peaks or dips. 
Extensions to non-integrable systems are discussed.

\end{abstract}


\maketitle

In both classical and quantum physics, the dynamics of a small system, such as a single oscillator, spin or molecule, is characterized by a handful of internal frequencies. 
These can be detected as resonances: driving the system with a periodic perturbation the dynamics at a resonance is characterized by an efficient transmission of energy to (and from) the system. 
More generally, the dynamics of a periodically driven system will result in a complex signal comprising beatings and revivals as well as full memory of the initial conditions~\cite{Hanggi_JCP97, Hanggi_book}.  
In turn, it is well known that dissipation may lead 
to a different scenario: in a single periodically-driven two-level quantum systems~\cite{Hausinger_PRA10, Russomanno_PRB11} a coupling, albeit weak, to an external environment leads at long times to the ``synchronization'' of the dynamics with the perturbation: 
its density matrix, after an initial transient, becomes time-periodic, and memory of the initial state is lost.


It is natural to ask whether the above scenario applies also to many-body, yet isolated, periodically driven quantum systems.
This was a purely academic question, until the recent progress in the context of ultra-cold atomic gases has made it possible 
to address experimentally their quantum dynamics \cite{Polkovnikov_RMP11,Bloch_RMP08}. 
If a many-body system is subject to a periodic driving, under which conditions is it going to get synchronized with the perturbation? 
And what kind of information will be kept on the initial state? 
Many-body systems, in the thermodynamic limit, are characterized by a continuum of frequencies, which suggests that the system absorbs indefinitely energy from the periodic driving~\cite{Polkovnikov_NatPhys11}. 
In addition, nonintegrable systems act as thermal bath for themselves~\cite{Polkovnikov_RMP11}; then one could expect a significant  loss of memory of the initial conditions, as indeed observed in a number of systems subject to an abrupt quench of a single parameter \cite{Rigol_Nat,SRED_PRE94,Kollath_PRL,Manmana_PRL}. 
%
While theoretical progress in non-equilibrium quantum many-body physics has been mainly concerned with  sudden quenches~\cite{Polkovnikov_RMP11} and slow annealings~\cite{Kadowaki_PRE98,Santoro_SCI02,Santoro_JPA06,Farhi_SCI01,Zurek_PRL05}, 
important theoretical~\cite{Eckardt_merd,Eckardt_PRL051,Kollath_PRL06,Kollath_PRA06,Mucco_JSM09,Arnab_PRB10,Arnab_PRB12,kollath_PRA11} and
experimental~\cite{Lignier_PRL07,Eckardt_PRA09,Sias_PRL08,Gemelke_PRL05,Zenesini_PRL09,doublon_PRL10} works dealing with the effects 
of a periodic driving on the coherent dynamics have recently appeared. 
In this paper we discuss the coherent dynamics of a periodically driven quantum many-body system. 
Using Floquet theory, we argue that, even if the dynamics is fully coherent, destructive interference in time makes
most of the observables attain, at long times, a time-periodic regime synchronous with the perturbation, 
in which memory of the initial state enters through some overlap factors. 
We explicitly demonstrate this for an integrable system, an inhomogenous quantum Ising chain subject to a time-periodic transverse field \cite{Arnab_PRB10,Arnab_PRB12,Mucco_JSM09},
where we show that the convergence towards a periodic steady regime follows whenever most of the Floquet spectrum is a continuum. 
We argue that this continuity condition is likely often met by many-body systems in the thermodynamic limit, perhaps
with the important exception of disordered systems, although a rigorous theory for general non-integrable systems is lacking. 
%
%
Focusing on the Ising chain, we discuss the dynamics of energy absorption, 
showing that the phase coherence of the system is seen transparently in the plethora of peaks or dips emerging in the frequency dependence of the various observables.  
Because of phase coherence, the system does not absorb energy indefinitely: 
the resulting density of excitations stays {\it well below} the maximum attainable value, 
as indeed observed in NMR systems \cite{Slichter:book}.
Consider a quantum many body system governed by a periodic Hamiltonian $H\left(t\right)=H\left(t+\tau\right)$ and the
quantum average $O(t)=\bra{\Psi(t)}\hat{O}(t)\ket{\Psi(t)}$ of an operator $\hat{O}(t)$ (possibly $\tau$-periodic),
$\ket{\Psi(t)}$ being the state of the system.
Thanks to Floquet theory~\cite{Hanggi_book,Shirley_PR65, Hausinger_PRA10}, 
we know that there is a complete basis of solutions of the Schr\"odinger equation (the Floquet states) which are periodic up to a phase 
$\ket{\Psi_\alpha(t)}=\nep^{-i\bar{\mu}_\alpha t}\ket{\Phi_\alpha(t)}$.
The many body quasi-energies $\bar{\mu}_\alpha$ are real, and the Floquet modes $\ket{\Phi_\alpha(t)}$ are periodic: 
both can be extracted from the knowledge of the evolution operator $\hat{U}(t,0)$ for $0\leq t\leq\tau$. 
Expanding $\ket{\Psi(t)}=\sum_\alpha R_\alpha\nep^{-i\bar{\mu}_\alpha t}\ket{\Phi_\alpha(t)}$ in the Floquet basis, with
the overlap factors $R_\alpha \equiv \langle {\Phi}_\alpha(0) | \Psi(0)\rangle$, we get:
\begin{equation} \label{Ot:eqn}
  O(t)=\sum_\alpha \left| R_\alpha \right|^2 O_{\alpha\alpha}(t) + \int_{-\infty}^\infty F_O(\Omega)\nep^{-i\Omega t}\ud\Omega
\end{equation}
where $O_{\alpha\beta}(t) = \bra{\Phi_\alpha(t)} \hat{O}(t) \ket{\Phi_\beta(t)}$ and
$F_O(\Omega) \equiv \sum_{\alpha\neq\beta} O_{\alpha\beta}(t) R_\alpha^*R_\beta \delta\left(\Omega-\bar{\mu}_\beta+\bar{\mu}_\alpha\right)$.
The first term, involving the diagonal elements of $O_{\alpha\beta}$, is clearly $\tau$-periodic, while 
the second term describes fluctuations due to off-diagonal elements, and has been recast as 
the Fourier transform of a (weighted) joint density of states with the quasi-energies in place of the usual energies.
If, in the thermodynamic limit, the Floquet spectrum approaches a continuum with the weights $O_{\alpha\beta}(t) R_\alpha^*R_\beta$ 
depending smoothly on the quasi-energies, then $F_O(\Omega)$ is a smooth function, whose Fourier transform
vanishes for large $t$ due to destructive interference in time (Riemann-Lesbesgue lemma), an internal dephasing
akin to inhomogenous broadening in NMR~\cite{Slichter:book}.
%
Memory of the initial state is, in principle, kept into 
the overlaps $R_\alpha$.

To exemplify our general argument, 
consider a quantum Ising chain: 
\begin{equation}  \label{hamil}
	\hat{H}(t) = -\frac{1}{2} \sum_{j=1}^{L} \left[ J_j  \sigma_j^z \sigma_{j+1}^z + h_j(t) \sigma_j^x \right]  \;.
\end{equation}
Here, the $\sigma^{x,z}_j$ are spins (Pauli matrices) at site $j$ of a chain of length $L$ with periodic boundary conditions $\sigma^{x,z}_{L+1}=\sigma^{x,z}_1$,  
the $J_j$ are longitudinal couplings, and the $h_j(t)$ transverse fields. 
%
This Hamiltonian can be transformed, through a Jordan-Wigner transformation \cite{Lieb_AP61}, to a ``solvable'' quadratic-fermion form. 
When the system is time-independent and homogenous, $h_j=h$ $J_j=J=1$, this model has two mutually dual gapped phases, a ferromagnetic ($|h|<1$), 
and a quantum paramagnetic ($|h|>1$), separated by a quantum phase transition at $h_c=1$.
In the general time-dependent and inhomogeneous case, the unitary dynamics of $H(t)$ can be studied through a time-dependent Bogoliubov {\em Ansatz},
see final discussion and supplementary material (SM).
Although much of what we will say applies also to the inhomogeous case, we start with a homogenous chain, $J_j=J=1$ and $h_j(t)=h(t)$. 
For a homogenous chain, going to $k$-space, $\hat{H}(t)$ becomes a sum of two-level systems: 
$\hat{H}(t)=\sum_k^{\rm ABC} \hat{H}_k(t) = 
\sum_{k}^{\rm ABC} \left[ \epsilon_k(t)  \left(c_k^\dagger c_k-c_{-k} c_{-k}^\dagger\right) - i\Delta_k \left(c_k^\dagger c_{-k}^\dagger-c_{-k} c_k\right)\right]$,
where $\epsilon_k(t) = h(t)-\cos{k}$, $\Delta_k=\sin{k}$, and the sum over $k$ is restricted to positive $k$'s of the form $k=(2n+1)\pi/L$ with $n=0,\ldots,L/2-1$,
corresponding to anti-periodic boundary conditions (ABC) for the fermions \cite{Lieb_AP61}, as appropriate for $L$ multiple of $4$, as we assume.
Each $\hat{H}_k(t)$ acts on a 2-dim Hilbert space generated by $\{ c_k^\dagger c_{-k}^\dagger \ket{0}, \ket{0} \}$, and can be represented in that basis by a $2\times 2$ matrix
$H_k(t)=\epsilon_k(t) \sigma^z + \Delta_k \sigma^y$, with instantaneous eigenvalues 
$\pm E_k(t)=\pm\sqrt{\epsilon_k^2(t)+\Delta_k^2}$.
%
%
%
Since $\hat{H}(t)$ conserves the fermion parity and the momentum, but mixes $\ket{0}$ with $c_k^\dagger c_{-k}^\dagger\ket{0} $, the state has
a BCS-like form $\ket{\Psi(t)}=\prod_{k>0}^{\textrm{ABC}}\ket{\psi_k(t)} = \prod_{k>0}^{\textrm{ABC}}\left(v_k(t)+u_k(t)c_k^\dagger c_{-k}^\dagger\right)\ket{0}$
with $i\hbar \left( \begin{array}{c} \dot{u}_k\\ \dot{v}_k \end{array} \right) = H_k(t) \left( \begin{array}{c} {u}_k\\ {v}_k \end{array} \right)$
(Bogoliubov-de Gennes equations). 
%
%

\begin{figure}
\begin{center}
   \includegraphics[width=8.2cm]{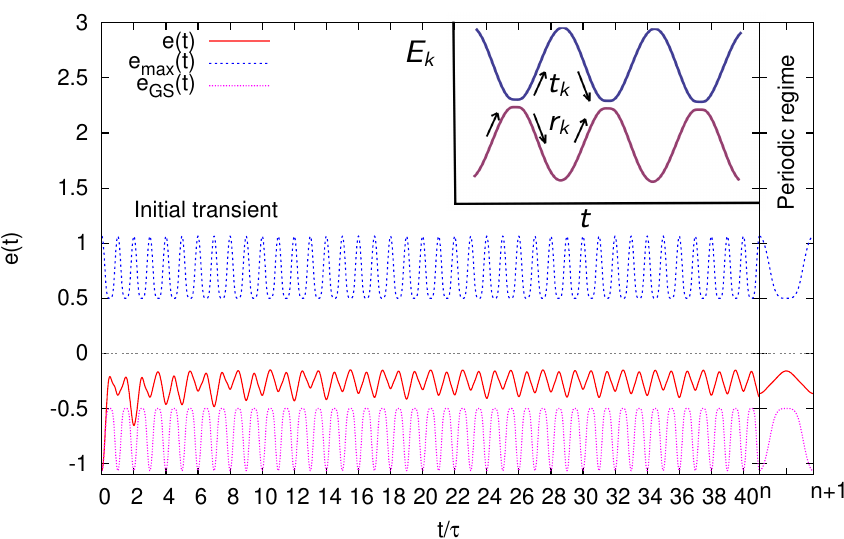}
\end{center}
\caption{
(Upper inset) Sketch of the dynamics within each $k$ subspace: 
At each avoided crossing the system can be either reflected by the gap or transmitted with amplitude $r_k$ and $t_k$. 
%
(Main figure) Evolution of the average energy density $e(t)= \bra{\Psi(t)} \hat{H}(t) \ket{\Psi(t)}/L$ 
versus $t$ with a driving field $h(t)=1+\cos{(\omega_0 t)}$, for $\hbar\omega_0/J=2$.
The lower and upper curves are the instantaneous ground-state and maximum energy density, versus $t$. 
Notice the initial transient (left panel, obtained from numerical solution with $L=2000$),
and the final periodic behaviour of $e(t)$ 
(right panel, obtained from $e^{\rm per}(t)$, again with $L=2000$).
}
\label{e_vs_t:fig}
\end{figure}

Assuming the simplest periodic modulation across the critical point, $h(t)=h_c+\delta h(t)=1+A\cos\left(\omega_0 t +\varphi_0 \right)$, 
the dynamics in time of each two-level system appears to be analogous to a sequence of Mach-Zehnder interferometers in space, see Fig.\ref{e_vs_t:fig}, 
each avoided crossing being analogous to a beam splitter where the system, starting in the ground state, is reflected or transmitted with an amplitude $r_k$ and $t_k$.
%
%
%
The phase accumulated in between two avoided crossings is crucial for the long-time dynamics: 
destroying phase coherence by repeated ``measurements'' of the energy at the end of each period, 
the system is reflected of transmitted with {\em probabilities}  $|r_k|^2$ and $|t_k|^2$, and one can easily 
prove \cite{Arnab_PRB10} that it would absorb energy indefinitely, ending up in an infinite temperature 
mixed state with equal weights $1/2$. 

Turning to the exact coherent evolution of our system, 
and applying Floquet theory~\cite{Hanggi_book,Shirley_PR65, Hausinger_PRA10} we (numerically) calculate, for each $k$, the
quasi-energies  $\mu_k^+=-\mu_k^-=\mu_k$ and the Floquet modes  $\ket{\phi_k^+(t)}$ in terms of which
the $k$-component of the state $\ket{\Psi(t)}$ reads: 
%
$\ket{\psi_k(t)} = r_k^+ \, \nep^{-i\mu_k\,t} \, \ket{\phi_k^+(t)}  \,+\, r_k^-  \,\nep^{i\mu_k\,t} \,\ket{\phi_k^-(t)} $
%
with $r_k^{\pm}=\bra{\phi_k^\pm(0)} \left.{\psi_k(0)}\right\rangle$. 
The continuous Floquet spectrum leads to the asymptotic decay of the fluctuations around the periodic steady-state regime.
Consider, for instance, the average energy $e(t)= \bra{\Psi(t)} \hat{H}(t) \ket{\Psi(t)}/L$, which
(as seen in Fig. \ref{e_vs_t:fig}) stabilizes well below the infinite-temperature value ($e=0$) into a well-defined periodic function 
$e^{(\rm per)}(t) =  \sum_{\alpha=\pm} \int_0^\pi \! \frac{ \ud k}{2\pi} \left| r_k^\alpha \right|^2 \bra{\phi_k^\alpha(t)} \hat{H}_k(t) \ket{\phi_k^\alpha(t)},\;{\rm after}$ 
a transient given by a Fourier-integral $\int_0^\pi \! \frac{ \ud k}{\pi}  \Re\left[{r_k^-}^*\,r_k^+\,\bra{\phi_k^-(t)}\hat{H}_k(t)\ket{\phi_k^+(t)} \,\nep^{-2i\mu_kt} \right]\;{\rm oscillating}$
and {\em decaying} to $0$ as a power-law when $t\to \infty$.  
A decaying transient for the transverse magnetization has been analyzed in Ref.~\cite{Arnab_PRB12}. 

\begin{figure}
\begin{center}
    \includegraphics[width=8.5cm]{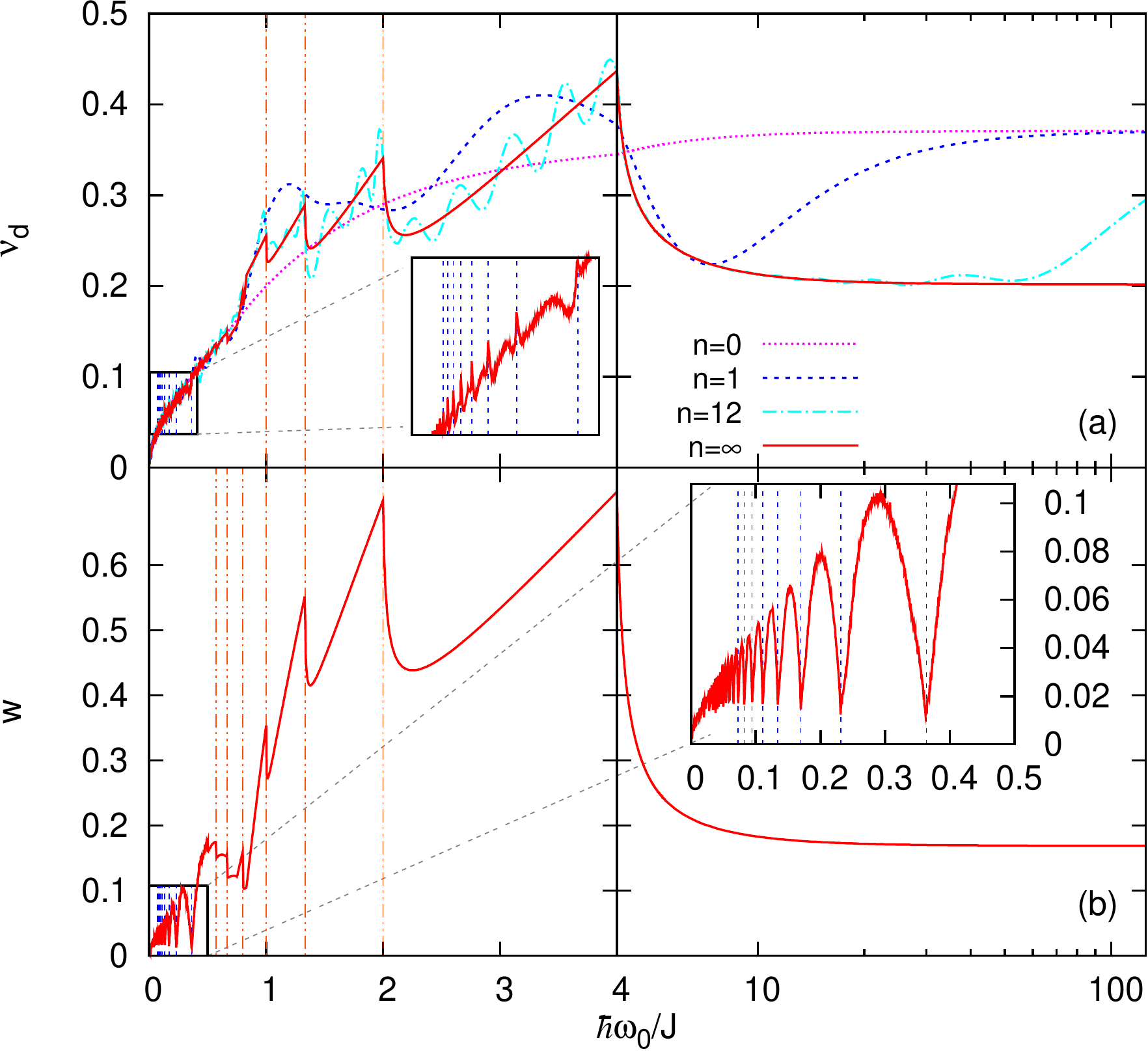}
\end{center}
\caption{(a) The density of defects for a driving field $h(t)=1+\cos{(\omega_0 t)}$ at the end of half-periods, i.e., 
at times $t=t_{2n+1}=(2n+1)\pi/\omega_0$ when the transverse field vanishes; $n=0$ is the result of a single LZ crossing.
(b) The total work (per spin) done on the system $w=e(t_{2n}\to \infty)-e(0)$ versus the frequency $\omega_0$ of the transverse field.} 
\label{total_difetti_work-crop:fig}
\end{figure}
%
Many interesting quantities can be extracted from $e(t)$ at different times. 
For definiteness, consider the case when the field oscillates as $h(t)=1+\cos{(\omega_0 t)}$ (i.e., $A=1$ and $\varphi_0=0$), 
making excursions between $2$ and $0$, repeatedly crossing the critical value $h_c=1$. 
$h$ vanishes for all $t=t_{2n+1}=(2n+1)\pi/\omega_0$, where 
the density of defects (kinks) $\nu_d$ of the classical Ising chain \cite{Zurek_PRL05} is 
$\nu_d(\omega_0,n) = \bra{\Psi(t_{2n+1})} \sum_{j=1}^{L} \left[ 1-\sigma_j^z \sigma_{j+1}^z\right] \ket{\Psi(t_{2n+1})}/(2L)=e(t_{2n+1})-e_{\rm GS}(t_{2n+1})$.
%
%
The index $n$ counts the number of times, $2n+1$, the system goes through the critical point.  
Fig.~\ref{total_difetti_work-crop:fig}a shows $\nu_d$ vs $\omega_0$ for $n=0,1,12,\infty$. 
After only one crossing ($n=0$) a single Landau-Zener (LZ) event occurs for the low-k critical modes:
the resulting curve resembles that for linear annealing of the transverse field \cite{Zurek_PRL05,Mucco_JSM09}:
$\nu_d\propto \sqrt{\omega_0}$ for $\omega_0\to 0$ (Kibble-Zurek scenario). 
The saturation for $\omega_0\to \infty$ is reproduced by the impulsive limit  
$\nu_d^{\rm imp}=\bra{\Psi(0)}  \sum_{j=1}^{L} \left[ 1-\sigma_j^z \sigma_{j+1}^z\right] \ket{\Psi(0)}/2L$, 
where the system remains frozen in the initial state $\ket{\Psi(0)}$.
For $n=1$, interference effects between the three LZ crossing encountered lead to peaks 
(at intermediate frequencies) and to a strong decrease of $\nu_d$ for larger frequencies. 
This tendency persists for larger $n$'s, until the impulsive limit is substituted by a new large-$\omega_0$ 
plateau with a markedly reduced $\nu_d$, consistently with the vanishing of the transient component of $e(t)$. 
In the limit $n=\infty$, interference effects between successive LZ crossings give rise to 
two series of peaks which occur for $\hbar\omega_0/J=4/p$, with $p=1,2,\dots$ and for the $\omega_0$'s for which $J_0(2J/\hbar\omega_0)=0$,
$J_0$ being the Bessel function of order $0$. 
Both series are easy to explain (see SM for a detailed discussion): 
the first  one is due to particular resonances at $k=\pi$, occurring in the spectrum of the Floquet quasi-energies $\mu_k$, 
and originating from multi-photon processes in the Shirley-Floquet Hamiltonian \cite{Shirley_PR65}; 
the second one originates from the behaviour of the critical modes with $k\to 0$ for which the rotating wave approximation 
\cite{Kayanuma_PRA94,Hausinger_PRA10,Arnab_PRB10} applies.  

\begin{figure}
\begin{center}
   \includegraphics[width=8.5cm]{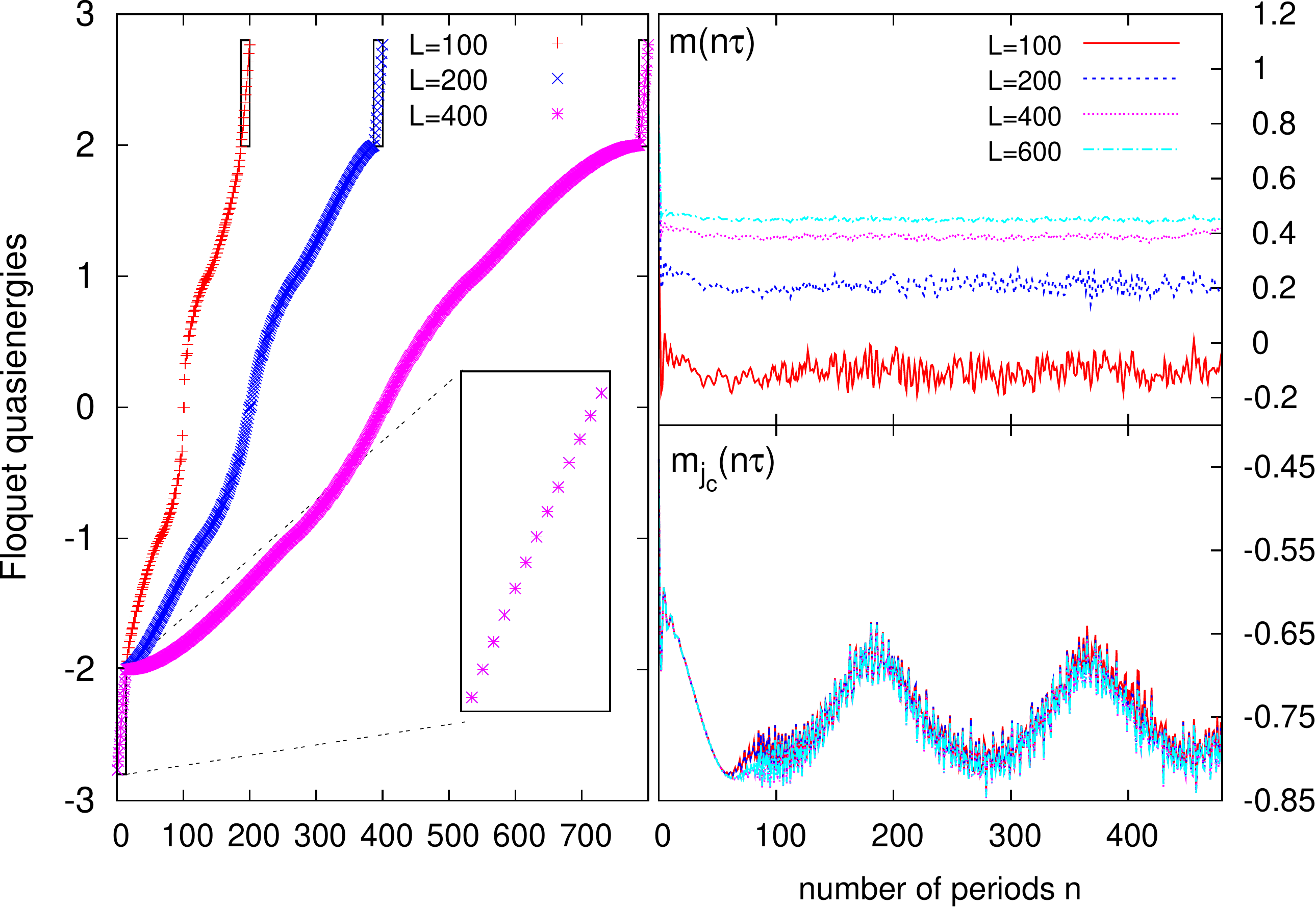}
\end{center}
\caption{(Left panel) The Floquet spectrum for an Ising chain with a Gaussian inhomogeneity with $l=20$, $h_G=2.8$, $\omega_0=10$, $A=1$ and different values of $L$.
Notice a finite $L$-independent number of discrete quasi-energies.  
(Upper right panel) The average transverse magnetization $m(t)$ probed at $t=n\tau$, showing fluctuations that decrease for increasing $L$.
(Lower right panel) The transverse magnetization at the center of the inhomogeneity, $m_{j_c}(t=n\tau)$, whose fluctuations persist for all $L$.}
\label{floquet:fig}
\end{figure}

A similar interference scenario occurs when we consider the work done on the system at the end of each period. 
At $t=t_{2n}=2n\pi/\omega_0$ the field has performed $n$ full periods of oscillation, and the work done (per spin) is $w_n=e(t_{2n})-e(0)$.
Fig.~\ref{total_difetti_work-crop:fig}b shows the results for the total work $w=w_{n\to \infty}$, as a function of the frequency $\omega_0$.
Notice that the total work is {\em finite} since the system stops absorbing a net amount of energy during the initial transient, and then settles down into a periodic
dynamics, during each cycle of which no net work is performed.
We clearly observe three different regimes for $w(\omega_0)$: i) a large-$\omega_0$ plateau, ii) an intermediate frequency region with peaks at $\hbar\omega_0/J=4/p$, 
iii) and a low-frequency region $w\sim \sqrt{\omega_0}$ with {\em dips} at the zeroes of $J_0(2J/\hbar\omega_0)$. 
In the SM we discuss also the average transverse magnetization $m(t)=\bra{\Psi(t)}\sum_j\sigma_j^x\ket{\Psi(t)}/L$.
The model discussed was very special: integrable, translationally invariant, and reducing (in $k$-space) to an assembly of two-level systems. 
Similar physics apparently emerges also in a periodically modulated homogeneous one-dimensional Hubbard model (Bethe-Ansatz-integrable at equilibrium),
as numerically found in Fig. 1 of Ref.~\onlinecite{Kollath_PRA06}.
If we break integrability, retaining translational invariance, the dynamics is non-trivial, and a rigorous theory is lacking.
Nevertheless, one could argue that the Floquet spectrum ${\bar\mu_\alpha}$ is likely to remain continuous in the thermodynamic limit. 
(Indeed, the Floquet spectrum is continuous even for a single atom under an ac electric field: its static discrete energy levels always hybridize 
with the unbounded continuum via ``multi-photon'' processes~\cite{yajima_CMP82}.)
We therefore expect, again, a periodic steady regime with vanishing fluctuations. 
If we break translational invariance, retaining integrability to carry on the analysis, 
e.g., by $h_j(t) = 1 + h_G\,\nep^{-\left(j-j_c\right)^2/2{l}^2}+A \cos(\omega_0 t)$
(a Gaussian inhomogeneity of width $l$, sitting at the central site $j_c=L/2+1$, in the transverse field), then one
has to solve a system of $2L\times 2L$ Bogoliubov-de Gennes $\tau$-periodic equations 
\footnote{As shown by C. Bloch and A. Messiah, Nucl. Phys. {\bf 39}, 95 (1962), a unitary $2L\times 2L$ matrix of the form relevant here, 
see SM, can be canonically decomposed into $2\times 2$ blocks; but the transformation is here time-dependent, hence the $2L\times 2L$ 
problem does not reduce to an assembly of two-level systems.}, 
whose Floquet quasi-energies show only a finite number of discrete states, see left panel of Fig.~\ref{floquet:fig}.
%
%
The discrete quasi-energies lead to delta-functions in $F_O(\Omega)$, see Eq.~\eqref{Ot:eqn}, which do not merge into a smooth continuum. 
Nevertheless, for extensive operators, e.g., the average transverse magnetization $m(t)$ 
(upper right panel of Fig.~\ref{floquet:fig}), or for local operators coupling mostly continuum Floquet modes, 
the fluctuating contributions due to such a finite number of delta-functions in $F_O(\Omega)$ vanish for $L\to \infty$. 
On the contrary, fluctuations persist for local operators sensitive to the discrete states,
e.g., the transverse magnetization at the center of the inhomogeneity $m_{j_c}(t)=\bra{\Psi(t)}\sigma_{j_c}^x\ket{{\Psi(t)}}$ (see lower right panel of Fig.~\ref{floquet:fig}).
%
The case of  {\em disordered} systems, whose static $H$ has an important pure-point spectral region, clearly calls for further studies, particularly on the role of a possible mobility edge
in presence of periodic driving.

The fact that the system does not absorb energy up to the infinite temperature state is most likely a consequence of phase coherence and integrability: 
one would expect that breaking integrability, i.e., making quasi-particles scatter inelastically, 
should lead to heating up to an infinite temperature state. 
This expectation, natural in the thermodynamic limit, is not obviously realized in finite-size systems as a result of the fact that 
a chaotic spectrum (such as that of non-integrable systems) 
gives rise in many cases to localization in energy space. Therefore, this issue has to be clarified with further studies.

In conclusion, we studied the coherent evolution of quantum many-body systems under periodic driving. 
On the basis of results obtained for an integrable inhomogenous Ising chain, we have argued that, under the hypotesis of a continuous Floquet spectrum, 
a large class of averages of observables, after an initial transient, would tend to synchronize with the driving into a time-periodic ``steady-state''. 

\acknowledgements
We acknowledge discussions with G. Biroli, A. Das, R. Fazio, M. Fabrizio, J. Marino, G. Menegoz, P. Smacchia, E. Tosatti and S. Ziraldo. 
Research was supported by MIUR, through PRIN-20087NX9Y7, by SNSF, through SINERGIA Project CRSII2 136287\ 1, 
by the EU-Japan Project LEMSUPER, and by the EU FP7 under grant agreement n. 280555.


\end{document}